\documentclass[aip,jmp,amsmath,amssymb,preprint]{revtex4-1}
\usepackage{color}
\usepackage{graphicx}
\usepackage{dcolumn}
\usepackage{bm}
\usepackage[utf8]{inputenc}

\begin{document}

\title[]{Structured Back Gates for High-Mobility Two-Dimensional Electron Systems Using Oxygen Ion Implantation}

\author{M. Berl}  \email{mberl@phys.ethz.ch}

\author{L. Tiemann}%
\affiliation{Solid State Physics Laboratory, ETH Zürich, 8093 Zürich, Switzerland
}%

\author{W. Dietsche}
\affiliation{Solid State Physics Laboratory, ETH Zürich, 8093 Zürich, Switzerland
}%

\author{H. Karl}
\affiliation{Lehrstuhl für Experimentalphysik IV, Universität Augsburg, 86159 Augsburg, Germany
}%

\author{W. Wegscheider}
\affiliation{Solid State Physics Laboratory, ETH Zürich, 8093 Zürich, Switzerland
}%

\date{\today}

\begin{abstract}
We present a reliable method to obtain patterned back gates compatible with high mobility molecular beam epitaxy (MBE) via local oxygen ion implantation that suppresses the conductivity of an 80$\,$nm thick silicon doped GaAs epilayer. Our technique was optimized to circumvent several constraints of other gating and implantation methods. The ion-implanted surface remains atomically flat which allows unperturbed  epitaxial overgrowth. We demonstrate the practical application of this gating technique by using magneto-transport spectroscopy on a two-dimensional electron system (2DES) with a mobility exceeding 20$\,\times\,$10$^{6}\,$Vs/cm$^{2}$. The back gate was spatially separated from the Ohmic contacts of the 2DES, thus minimizing the probability for electrical shorts or leakage and permitting simple contacting schemes.
\end{abstract}


\maketitle

Reliable electrostatic gating is a key element in most experiments in semiconductor physics to tune the charge carrier density of a two-dimensional electron (2DES) or hole (2DHS) system, which is intrinsically defined by doping. For sophisticated mesoscopic systems or nanostructures, gates need to be patterned. While structured top gates can be obtained with relative ease during the sample fabrication, patterning back gates is demanding and each available method has certain technological limitations.\par

Structured back gates can be obtained by thinning down the substrate, followed by the evaporation of metal gates on the backside\cite{Eisenstein90,EBASE}. The fragility of these thinned samples, however, make their handling difficult. In addition, as a result of the large gate separation, substantial gate voltages have to be applied. Several alternative technologies have therefore been exploited to pattern back gates on the epitaxial side of the wafer. For overgrown prestructured back gates, the gates are wet-chemically etched out of a highly doped layer on the epitaxial side of the substrate\cite{SMOOTH,Tiemann08,Huang12,Rubel98}, prior to the growth of the heterostructure. Since high mobility epitaxy critically depends on the surface roughness, the etched surface of these systems will not favor high mobility 2DESs. A different approach is the implantation of oxygen ions to locally suppress the conductivity of a doped epilayer on an otherwise insulating wafer. It is a commonly used industrial technique and has also been applied in research scale molecular beam epitaxy \cite{IMPLANT,IMPLANT2}. However, along with focused ion beam implantation\cite{Brown94}, all these methods can result in an amorphization of the surface and are not tailored to highest mobility samples required in fundamental research.\par

Here we report on an optimized technique using oxygen ion implantation which circumvents the limitations of other available back gating technologies. Our method allows to structure entire wafers for high mobility heterostructure growth by using standard optical photo resist as a shadow mask for the impinging oxygen ions. Atomic force microscopy (AFM) confirms that the substrate surface remains atomically flat after the implantation which is the prerequisite for high mobility molecular beam epitaxy. We use standard magneto-transport measurements to demonstrate the suitability and reliability of this gating method and the compatibility with high mobility MBE.\par

\begin{figure}[t!]
\centering
		\includegraphics[width=\textwidth]{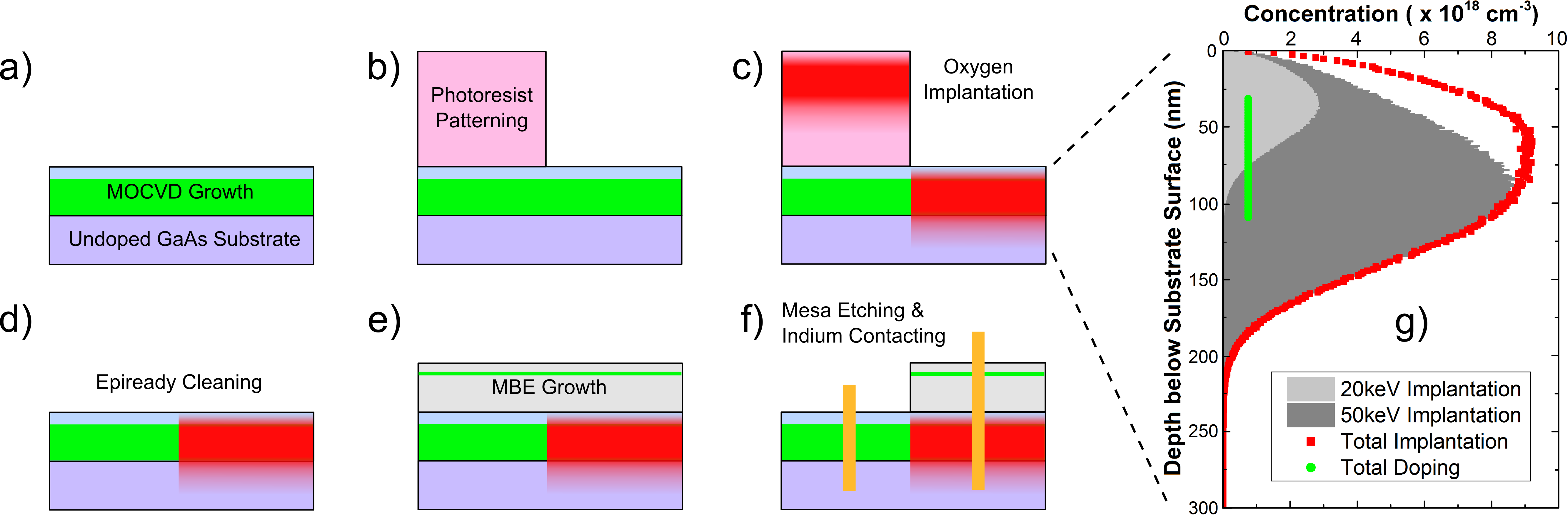}
\caption[Messdaten]{a)-f) Sample processing scheme; a) MOCVD growth of an 80$\,$nm Si doped GaAs epilayer (green layer) and a 30$\,$nm thick undoped GaAs cap grown on a (001) semi-insulating GaAs wafer b) Patterning with photo resist c) Oxygen ion implantation d) Epiready cleaning of the surface after photo resist removal e) MBE overgrowth including a quantum well confined 2DES (green layer) f) Wet chemical mesa etching and removal of the 2DES over the back gate contact areas g) Implantation depth and doping profile within the GaAs substrate. The total implantation (red squares) is the sum of two implantation distributions with 1.6$\,\times\,$10$^{13}\,$Ions/cm$^{2}$ at 20$\,$keV (light grey shaded area) and 1.0$\,\times\,$10$^{14}\,$Ions/cm$^{2}$ at 50$\,$keV (dark grey shaded), which were determined from TRIM simulations\cite{TRIM}. The doping layer (30$\,$nm-110$\,$nm) with a doping concentration of 7.5$\,\times\,$10$^{17}\,$cm$^{-3}$ is indicated by the green bar.}
	\label{fig:subband}
\end{figure}

The fabrication begins with a metal organic chemical vapor deposition (MOCVD) of a 80$\,$nm thick homogeneous silicon doped GaAs epilayer grown on a semi-insulating two inch GaAs wafer (Fig. 1a). The doping layer is covered by a 30$\,$nm thick undoped GaAs caping layer to ensure that the peak value of the subsequent implantation will be located within the doping layer. The shift of the peak further into the substrate also reduces the implantation damage to the surface. We experimented with doping concentrations ranging from 3$\,\times\,$10$^{16}\,$cm$^{-3}$ to 3$\,\times\,$10$^{18}\,$cm$^{-3}$ and found 7.5$\,\times\,$10$^{17}\,$cm$^{-3}$ to be the most reliable choice for the subsequent ion implantation. If the doping concentration is too high, it is not possible to suppress the conductivity by oxygen implantation\cite{REPORT}. If the doping concentration is too low, the resistivity of the back gate and its Ohmic contacts will be too high. After the MOCVD growth, the wafer is covered with photo resist (Microchemicals GmbH AZ1518) and patterned by standard optical photo lithography (Fig. 1b). The photo resist is sufficiently thick ($\sim\,$2$\,\mu$m) to work as a selective absorber, i.e., shadow mask, for the impinging oxygen ions so that only the exposed parts will become implanted with ions (Fig. 1c). The reliable operation of our gating technique crucially depends on the interplay between doping concentration and ion implantation profile which is fundamentally given by the dose and the energy of implantation. To ensure that most of the ion incorporation is targeted at the 80$\,$nm thick doping layer, we have chosen two implantation doses of different energies after determining the optimal implantation distribution with a TRIM simulation\cite{TRIM} (Fig. 1g). We tested several pairs of doses and found 1.6$\,\times\,$10$^{13}\,$Ions/cm$^{2}$ at 20$\,$keV and 1.0$\,\times\,$10$^{14}\,$Ions/cm$^{2}$ at 50$\,$keV to be the optimal dose which allows for a stable suppression of conductivity while minimizing the impact of implantation to the surface crystal structure. The atomically smoothness of the implanted surface was confirmed with AFM measurements. The measured root mean square roughness is smaller than 0.5$\,$nm, which is comparable to the roughness of non-implanted surfaces.\par

Following implantation, the photo resist has to be carefully and completely removed from the wafer with acetone and plasma ashing. The wafer is then undergoing an epi-ready treatment (Fig. 1d), which consists of a surface cleaning with H$_{2}$SO$_{4}$ acid and a surface passivation by controlled surface oxidation on a hotplate (3 minutes at 300$^\circ\text{C}$). The epi-ready process is necessary to prepare the surface for the MBE overgrowth. For the heterostructure to be overgrown we choose comparable conditions as used for standard high mobility structures, which includes an arsenic-free bake-out of the wafer at around 450$^\circ\text{C}$ and a subsequent high temperature treatment (630$^\circ\text{C}$ for about 5 hours) during the heterostructure growth (Fig. 1e). After this high temperature treatment we observed changes in the resistivity of the implanted layers. These changes, which are known to arise from the annealing of the implantation induced defects\cite{ANNEALING,Muessig}, however, are negligible and do not affect the gate operation.\par

\begin{figure}[t!]
\centering
		\includegraphics[width=0.7\textwidth]{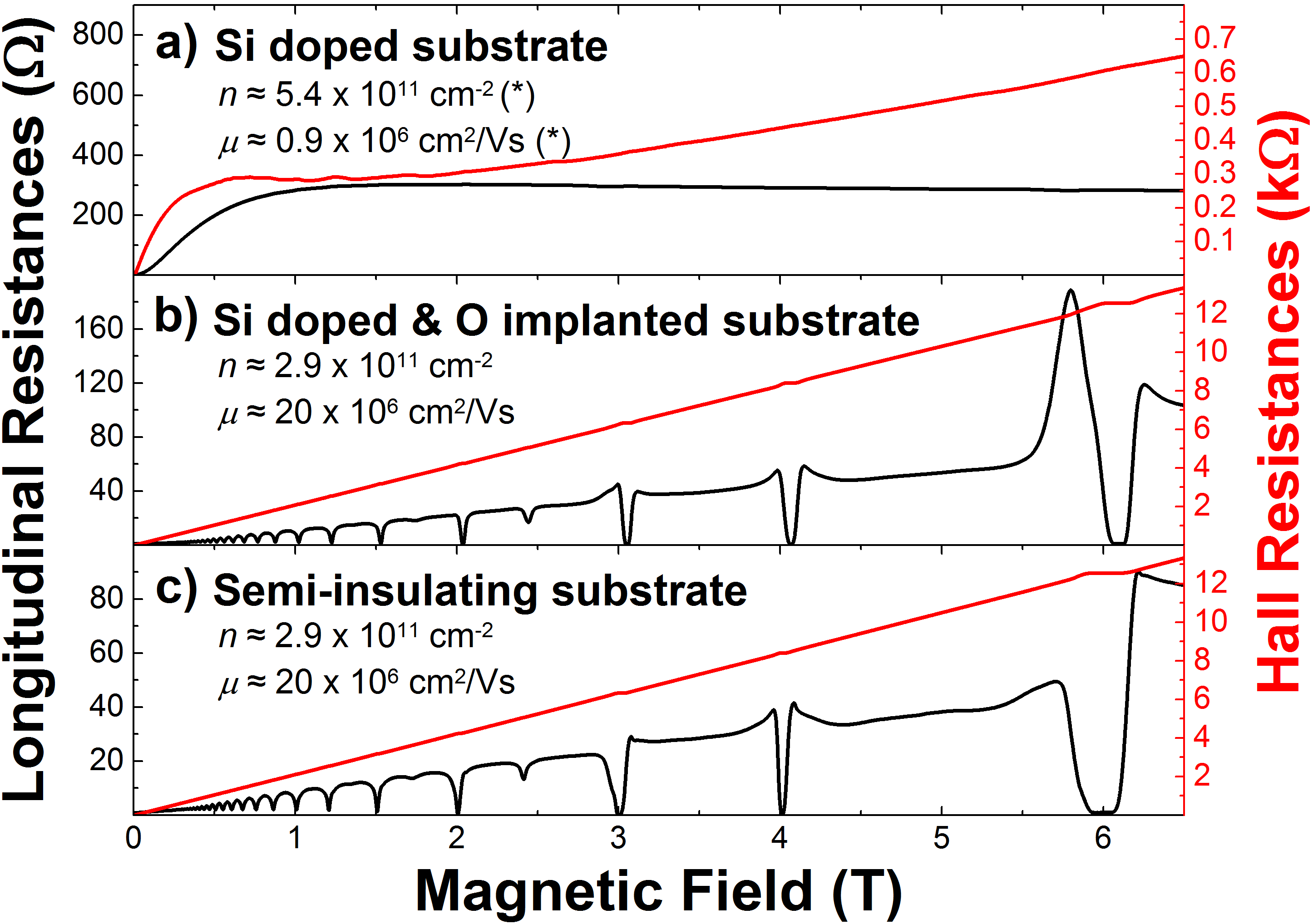}
\caption[Messdaten]{Magneto transport measurements at T$\,\approx\,$1$\,$K on van der Pauw samples with the same high mobility heterostructure (HS1) grown on different types of substrates; b) Substrate is silicon doped but not oxygen implanted; c) Substrate is silicon doped and oxygen implanted; d) Substrate is neither silicon doped nor oxygen implanted. (*) Values were calculated from Hall measurements at low magnetic field.
}
	\label{fig:subband}
\end{figure}

We have grown several GaAs/AlGaAs high mobility heterostructures on back gate patterned wafers and now present the results from two of these MBE overgrown heterostructures, labeled as HS1 \& HS2. HS1 is doubled-sided delta-doped, whereas HS2 is single-sided delta-doped on the surface-facing side of the quantum well to avoid screening effects between 2DES and gate. Both heterostructures have a 27$\,$nm wide GaAs quantum well buried 230$\,$nm below the sample surface. The setback between doping layers and 2DES is 80$\,$nm and the distance between back gate and 2DES is $\sim\,$1$\,\mu$m. The Al concentration was kept low ($\sim\,$24$\%$) around the 2DES to obtain high mobilities but was increased towards the substrate in several steps up to $\sim$88$\%$ to prevent gate leakages.\par

Standard lock-in magneto-transport measurements on square van der Pauw samples and soldered indium contacts were used to determine whether the insulating and conductive parts of the sample are working properly after the high mobility heterostructure overgrowth. At low temperatures and high (perpendicular) magnetic fields, the density of states of the 2DES condenses into a dispersionless bands of quantized Landau levels. As a result of this Landau quantization, the longitudinal resistance displays Shubnikov–de Haas oscillations whereas the transversal Hall resistance displays plateaus at integer values of the filling factor $\nu\propto \frac{n}{B}$; known as the integer quantum Hall effect \cite{Klitzing}. Figure 2a shows magneto-transport measurements on a van der Pauw sample which was grown on a conductive silicon doped substrate without implantation. In this sample the contacts penetrate through the heterostructure including the conductive substrate. The quantum Hall effects exhibited by the 2DES are therefore superimposed to the parasitic parallel conductance originating from the silicon doped substrate. However, when the implantation sufficiently suppresses the substrate conductance, only the pristine quantum Hall effects (Fig. 2b) of the 2DES are observed. Charge carrier mobilities of 20$\,$Mio.$\,$cm$^{2}$/Vs could be measured, demonstrating unambiguously the suitability of implanted substrates for the delicate high mobility heterostructure overgrowth. To study the impact of the implantation on  high mobility heterostructure growth, we have performed measurements on an identical heterostructure grown on a semi-insulating GaAs wafer (Fig. 2c) which was neither doped nor ion-implanted. The magneto-transport measurements clearly show that both the signatures of the quantum Hall effects and mobilities and carrier densities are nearly identical to that from the implanted substrate.\par

\begin{figure}[t!]
\centering
		\includegraphics[width=0.7\textwidth]{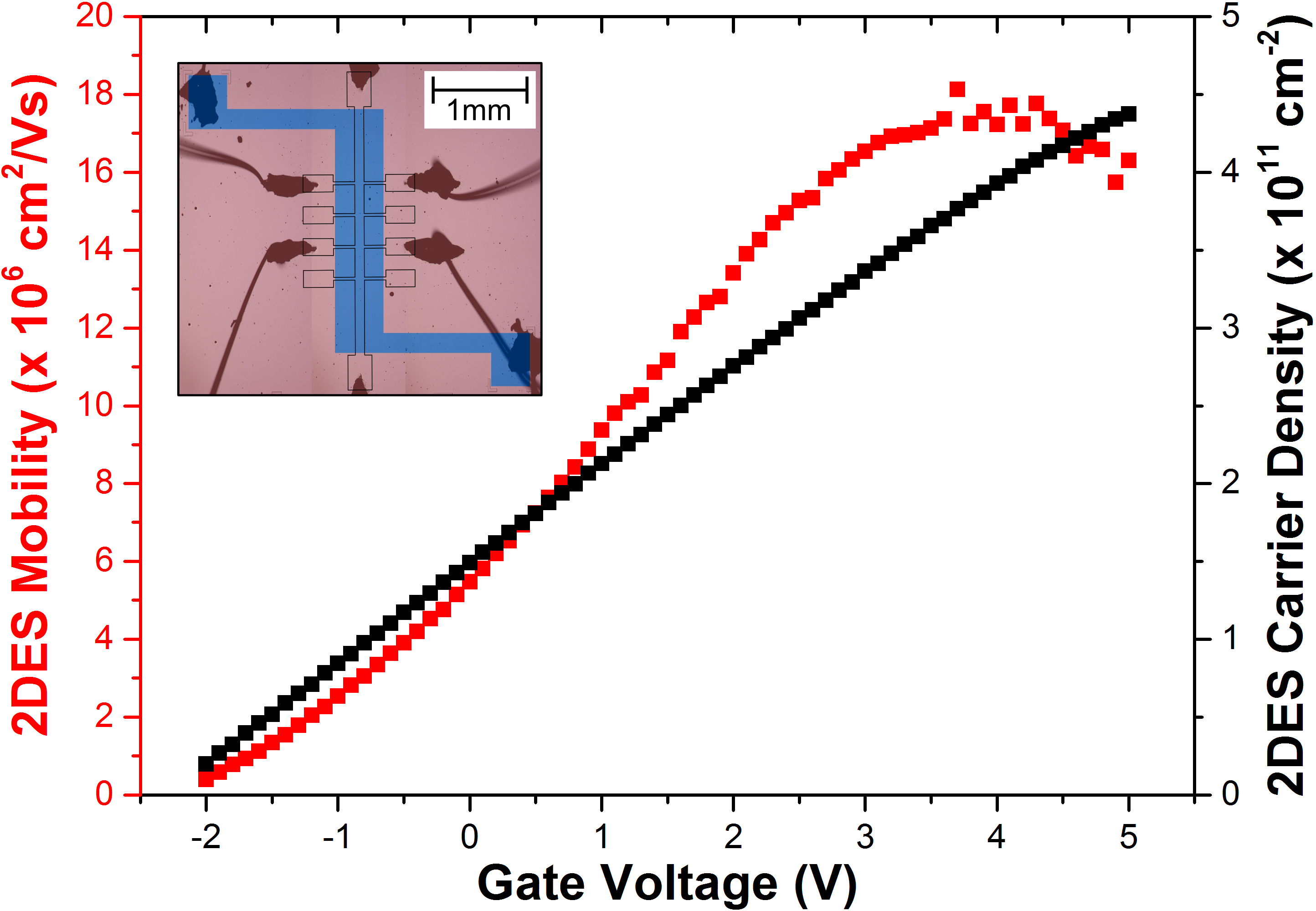}
\caption[Messdaten]{2DES mobility and carrier density as a function of the back gate voltage on a Hall bar structure using HS2 (T$\,\approx\,$25$\,$mK). Inset: Picture of an indium (black spots) contacted Hall bar mesa with the implanted substrate areas highlighted in red and the non-implanted conductive substrate parts highlighted in blue.}
	\label{fig:subband}
\end{figure}

High mobility heterostructures are the testbed for fragile exotic quantum effects. A prominent example are the fractional quantum Hall effects \cite{Tsui,Laughlin} that arise due to electron-electron interactions at non-integer values of the filling factor. To demonstrate the utilization and advantages of an implantation patterned back gate, we present low temperature measurements on standard Hall bar structures of high mobility heterostructures (HS1 \& HS2).\par

As depicted in the inset of Fig. 3, the inner Hall bar region of our Hall bar structure is located above the back gate, i.e., the region which was not ion-implanted, whereas the region underneath the contact arms was rendered non-conductive via oxygen ion bombardment. The exposed contacts (including the contact to the back gate) can therefore directly be contacted without risking an electrical short between back gate and 2DES. The only lithographic step which is necessary before making indium solder contacts is a wet-chemical etching process to define the Hall bar mesa and to remove the 2DES above the back gate contact areas (Fig. 1f). For HS2 the charge carrier density of the 2DES can be tuned over a wide range (Fig. 3) without provoking significant leakage current from the gate (I$_{leakage}\lesssim\,$30 pA). Over almost the entire range the carrier density is responding linearly to the gate voltage. With increasing density the  mobility could be raised from unbiased $\sim\,$6 Mio cm$^{2}$/Vs up to $\sim\,$18 Mio. cm$^{2}$/Vs by applying a gate voltage of about +3.5V.\par

\begin{figure}[t!]
\centering
		\includegraphics[width=0.7\textwidth]{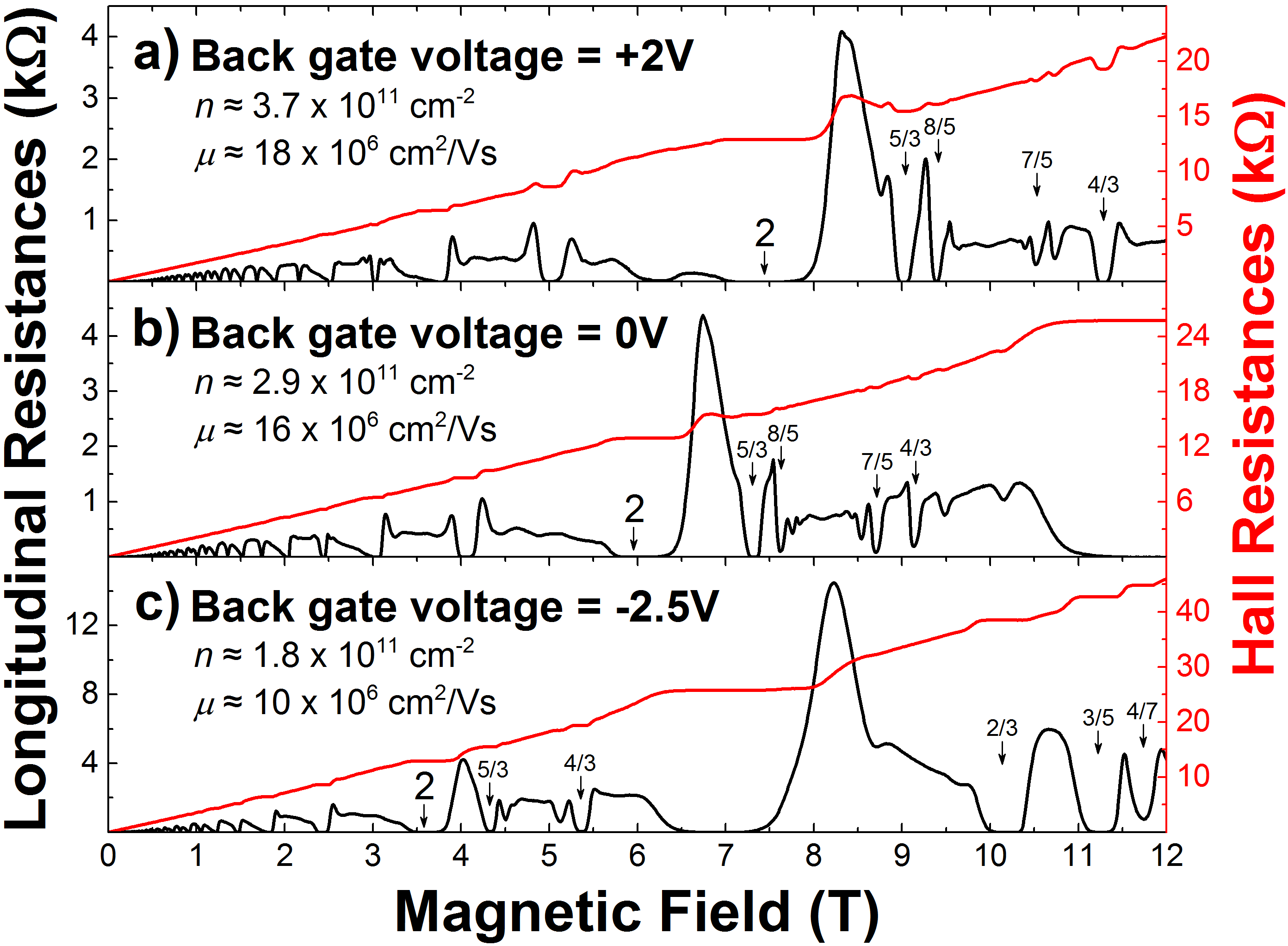}
\caption[Messdaten]{Magneto transport measurements at T$\,\approx\,$25$\,$mK on a Hall bar structure using HS1 with gate voltages applied from -2.5$\,$V to +2$\,$V. As a guide to the eye the Landau level filling factor 2 is labeled in each measurement. For reference, some fractional quantum Hall states were labeled as well.}
	\label{fig:subband}
\end{figure}

For HS1 low temperature magneto-transport measurements on the high mobility heterostructure are shown in Fig.4 for three exemplary gate voltages. The existence of a variety of fractional quantum states are indicative of the very high sample quality. Coincident with density changes, the observed fractional quantum Hall states are becoming more and less distinct. The highest mobility ($\sim\,$18$\,$Mio.$\,$cm$^{2}$/Vs) was achieved by increasing the charge carrier density to $\sim\,$3.7$\,\times\,$10$^{11}\,$cm$^{-2}$ via the back gate.\par

In summary, we demonstrated that the conductivity of a 80$\,$nm thick silicon doped GaAs-substrate can be suppressed by oxygen implantation. No negative influences to the sample surface, the heterostructure growth or the overall quality of the high mobility 2DES were observed. The functionality of a patterned ion-implantation to define back gate structures was demonstrated with magneto-transport measurements. The method we presented here is very reliable and gate-able structures can be produced with relative ease. We believe that this method will also be suitable for a variety of nanostructures, which currently rely on top gating schemes, as well as for an improved concept for double layer 2DES/2DHS systems\cite{Brown94,Rubel98,Tiemann08}.\par

The authors acknowledge financial support by the Swiss National Science Foundation (SNF) and the NCCR QSIT (National Competence Center in Research - Quantum Science and Technology. We thank Emilio Gini (ETH-FIRST cleanroom staff) for providing us with n-doped MOCVD substrates.

\end{document}